\documentclass[prb,amsmath,amssymb,
reprint,%
]{revtex4-2}

\usepackage{graphicx}
\usepackage{dcolumn}
\usepackage{bm}

\usepackage[utf8]{inputenc}
\usepackage[T1]{fontenc}
\usepackage{mathptmx}
\usepackage{xcolor}

\usepackage[colorlinks=true,linkcolor=blue,citecolor=blue,urlcolor=blue,allbordercolors=white]{hyperref} 

\begin{document}

\title{Mechanistic models of cell-fate transitions from single-cell data}

\author{Gabriel Torregrosa}
\email{gabriel.torregrosa@upf.edu}
\affiliation{Department of Experimental and Health Sciences, Universitat Pompeu Fabra,
Barcelona Biomedical Research Park, Dr. Aiguader 88, 08003 Barcelona, Spain}

\author{Jordi Garcia-Ojalvo}
\email{jordi.g.ojalvo@upf.edu}
\affiliation{Department of Experimental and Health Sciences, Universitat Pompeu Fabra,
Barcelona Biomedical Research Park, Dr. Aiguader 88, 08003 Barcelona, Spain}

\begin{abstract}
Our knowledge of how individual cells self-organize to form complex multicellular systems is being revolutionized by a data outburst, coming from high-throughput experimental breakthroughs such as single-cell RNA sequencing and spatially resolved single-molecule FISH.
This information is starting to be leveraged by machine learning approaches that are helping us establish a census and timeline of cell types in developing organisms, shedding light on how biochemistry regulates cell-fate decisions.
In parallel, imaging tools such as light-sheet microscopy are revealing how cells self-assemble in space and time as the organism forms, thereby elucidating the role of cell mechanics in development.
Here we argue that mathematical modeling can bring together these two perspectives, by enabling us to test hypotheses about specific mechanisms, which can be further validated experimentally.
We review the recent literature on this subject, focusing on representative examples that use modeling to better understand how single-cell behavior shapes multicellular organisms.
\end{abstract}

\maketitle

\section{Introduction}

Single-cell technologies are producing large amounts of data about multicellular development, which have created an urge to develop tools to extract information from those datasets.
In that respect, machine learning has proven to be very useful in many applications, where it has exhibited substantial predictive power. This raises the question of why bother trying to make mechanistic models.
However, as it has already been debated \cite{huang2018tension,del2020importance}, mechanistic and machine-learning approaches are not mutually exclusive.
The power of machine learning is the ability to find patterns in vast amounts of data in an inductive way. On the other hand, mechanistic models extend the deductive capacity of human reasoning, allowing us to apply engineering design methods \cite{santos2020multistable, lormeau2021rationally}.
Both perspectives can complement each other to make better theories and predictions \cite{baker2018mechanistic}.
This mutual interaction has been present for decades: the first (mechanistic) neuron model \cite{McCulloch:1943aa} served as inspiration for the generation of the network approach that has culminated in the development of deep neural networks.
There is renewed interest in this synergistic point of view, as evidenced by the efforts towards the development of tools  ranging from the selection of simpler and more informative models \cite{proulx2017untangling} to the analysis of complex systems using machine-learning strategies \cite{wang2019massive}, and including the proposal of benchmarks for modelling and data analysis \cite{veleslavov2020repeated,pratapa2020benchmarking}.

\section{Experimental single-cell breakthroughs}

Two main breakthroughs are providing an unprecedented push to study the impact of fate decisions at the single-cell level on the development of multicellular organisms.
The first one is the fast development of high-throughput techniques to measure gene expression at single-cell resolution, which are leading to new insights into the cell-fate transitions underlying multicellular development  \cite{briggs2018dynamics,Nowotschin:2019aa}.
The data is increasing in detail, as shown by the addition of key information such as spatial resolution \cite{vickovic2019high, rodriques2019slide, eng2019transcriptome}.
While these techniques are not perfect and many challenges are still open \cite{lahnemann2020eleven}, datasets continue to improve in richness and quality, offering promising avenues of research in developmental biology. 
The second breakthrough is the establishment of gastruloids as novel experimental models of development \cite{Brink:2014aa}.
These developmental models have shown the ability to reproduce with unprecedented detail the main features of embryo formation, from 3D organization \cite{beccari2018multi} to somitogenesis \cite{van2020single}.
The benefits of these systems are their increased experimental controllability, as well as the scalability of the number of samples that can be analyzed under different perturbations. 

\section{Building mechanistic models at the single-cell level}

In spite of the increase in the amount of data available, mechanistic models suffer from what has been called the ``curse of parameter space'' \cite{tyson2020dynamical}, namely a lack of knowledge of the \emph{in vivo} parameter values of most biochemical and mechanical processes within live cells.
This problem can be circumvented by (i) using dynamical behavior as a constraint of the models \cite{Kirk:2008ix}, (ii) applying the tools of dynamical systems (in particular bifurcation theory) to explore systematically how the different biological processes impact qualitatively the behavior of the system in both model and experiments \cite{Espinar:2013fe}, and (iii) using minimal models to describe the spatiotemporal self-organization of tissues
 \cite{hiscock2015mathematically,schweisguth2019self}.
In what follows, we describe several case examples illustrating the use of those methods, in particular of minimal models that can capture the fundamental principles of the cell-fate transitions that underlie the formation of organisms.

\subsection{Geometric description of cell-fate dynamics}

\begin{figure}[htb]
	\centering
	\centerline{\includegraphics[scale=.2]{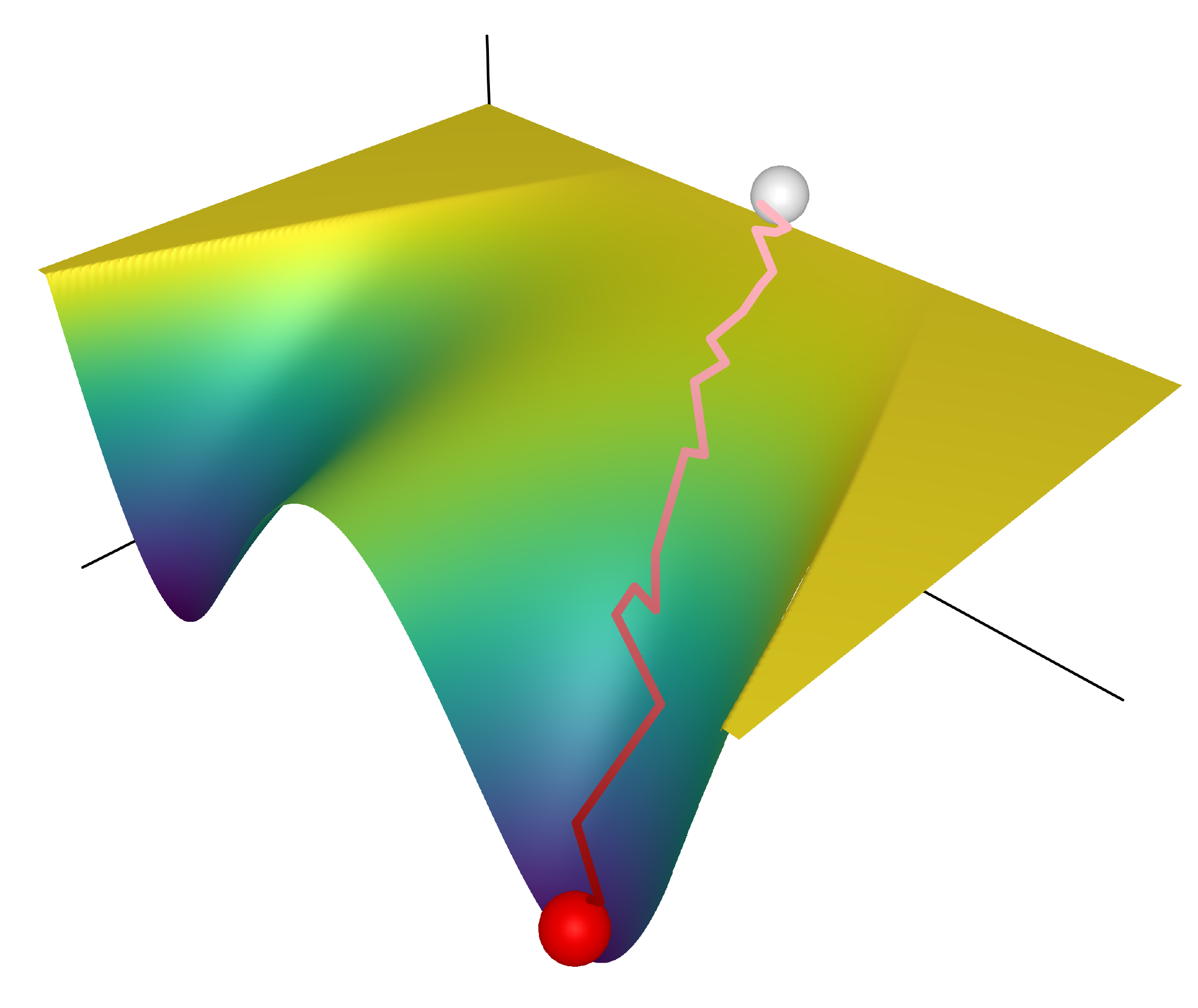}}
	\caption{A cell undergoing a stochastic trajectory down Waddington's landscape.}
	\label{fig:0}
\end{figure}

\begin{figure}[htb]
	\centering
	\centerline{\includegraphics[width=.4\textwidth]{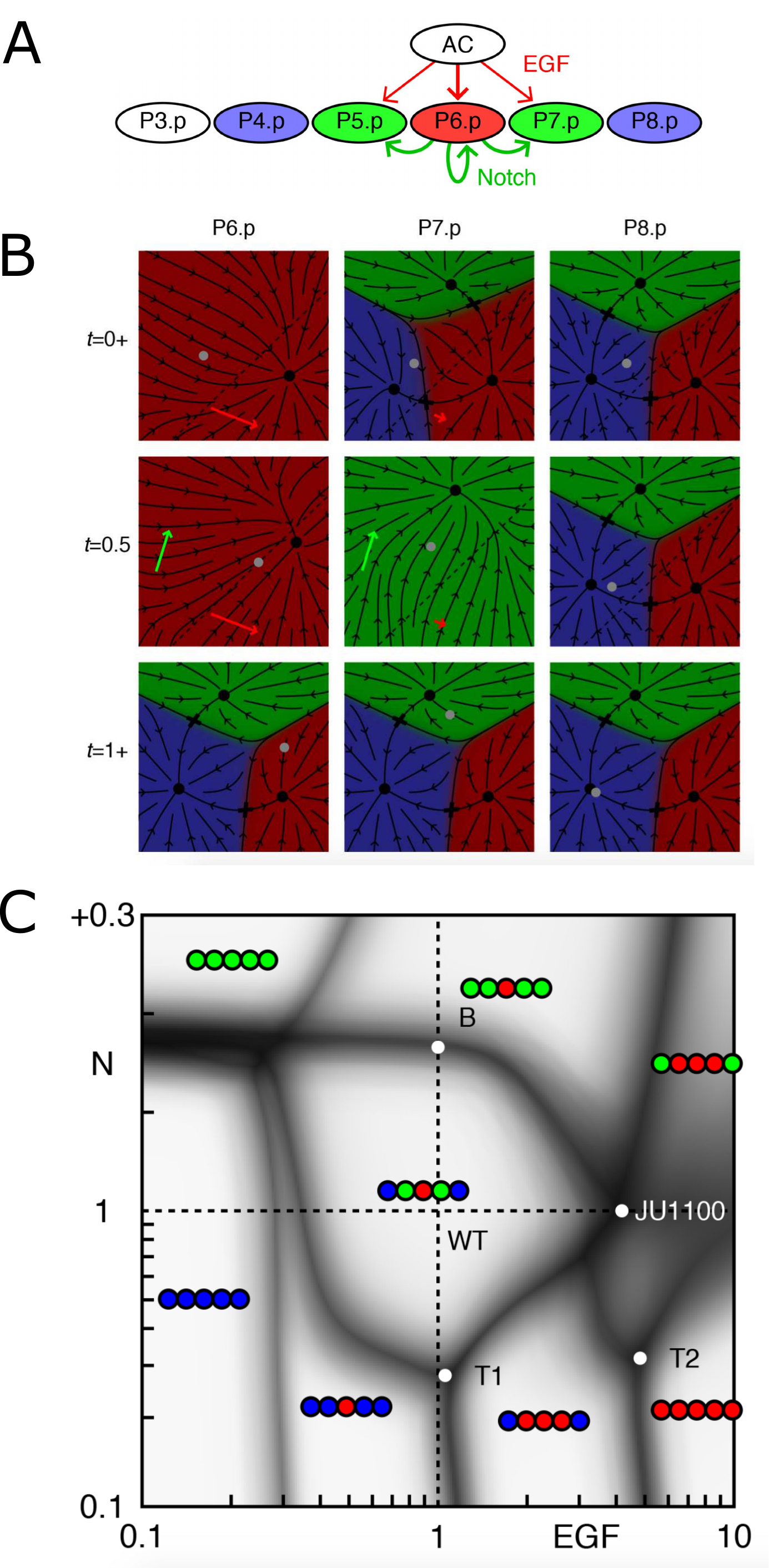}}
	\caption{Geometric approach to cell-fate transitions. 
	A: The vulval precursor cells (VPCs) in \emph{C. elegans} take fates based on the positional EGF cue from the anchor cell (AC, red arrows).
	Additionally, VPCs communicate with each other through the Notch pathway (green arrows).
	B: Both the EGF cue and the cell-cell communication modulate the landscape felt by each cell according to their position (left to right) and across time (top to bottom), reaching an attractor by the time the process finishes (at time $t=1$).
	C: Phase diagram showing the final outcome of the fate decision depending on the strengths of the Notch and EGF pathways.
	Tricritical points (T1 and T2) are non-trivial outcomes of the complex interaction between cells.
	Adapted from \cite{corson2017gene}.}
	\label{fig:1}
\end{figure}

The simplest paradigm of organism development comes from the epigenetic landscape picture introduced by Waddington, in which cells start from a pluripotent state and roll down through developmental valleys towards one of several alternative committed states (Fig.~\ref{fig:0}) \cite{Minelli:2014aa}.
This qualitative description is amenable to be interpreted with the formalism of dynamical systems \cite{Ferrell:2012qr,moris2016transition}.
Starting from this conceptual ``landscape model'', Corson and Siggia used the geometrical perspective of dynamical systems to explain the characteristic fate patterning of vulval development in \emph{C. elegans} \cite{corson2017gene}.
Fate commitment in these cells involves both positional EGF cues from an anchor cell and lateral cell-cell communication among the vulval precursor cells via Notch (Fig.~\ref{fig:1}A).
To understand how the system integrates the two types of signals, the authors ignored all biochemical details of the system, focusing on the fate patterning across a range of conditions.
With this information, they created the simplest model that captures the shape of the landscape over which the cells evolve towards the committed fates (Fig.~\ref{fig:1}B).
Using the effective degrees of freedom of the system to direct behaviors of the data allowed the authors to explore systematically the properties of the model, and study in particular the epistatic interactions between the exogenous EGF cue and the Notch-based cell-cell communication (Fig.~\ref{fig:1}C).

From the low complexity of the model, which consists only on eight fixed cells with two degrees of freedom each that represent the effective interactions between them, we can extract several lessons.
First, the model overcomes the problem of lack of full understanding of the vast number of biochemical and biophysical interactions underlying the fate-decision process, by focusing on an effective scale relevant to the epigenetic regulation of this process. This shows the strength of modeling in working constructively over different levels of complexity. The model exploits the formal and well understood description of dynamical systems to suggest alternative interpretations to our current understanding of the biology of the system, and to propose insightful experiments to test our current state of knowledge \cite{corson2017gene, camacho2021quantifying}.
The second lesson is the ability of this approach to discern among one of several theories, usually considered mutually exclusive, when interpreting experimental observations.
In this case, the model captures the joint effect that extrinsic cues and mutual cell-cell communication have on the system, conciliating an apparent contradiction between the two mechanisms.
The simultaneous dependence on diverse factors is a natural assumption given the complexity of biology.
Other studies have reached similar conclusions for phenomena ranging from the refinement of cell-fate patterning \cite{corson2017self} to robustness inactivating mechanisms \cite{gross2019guiding, green2015positional}.
Being able to select or integrate multiple models points at the benefits that single-cell resolution provides.

\subsection{A single-cell dynamical landscape}

From a cell-centric perspective, the genome contains all the necessary information describing the cell-fate landscape. However, cells are not isolated systems.
They feel external stimuli and gradients, and are sensitive to temperature and to mechanical interactions, both with the environment and between them.
All these factors impact the individual cells, creating complex landscapes.
In principle, we could imagine the global landscape of the full dynamical system containing all its degrees of freedom, both mechanical and biochemical, for all cells.
However, this picture is not commonly used for several reasons.
First, such an image obscures the elegant geometric understanding that we have of dynamical systems, and hence prevents us from extracting any possible intuition from it.
Second, and more practically, individual cells inside an organism are amenable to be interpreted as identical cell blocks, conferring the complex collective space with a valuable assumption of symmetry.

Considering each cell's dynamical system as a building block of the population, we find two common approaches to study cell-fate decisions at the collective level: continuous models (CM) and agent-based models (ABM).
Continuous models exist since the inception of the mathematical study of morphogenesis with the pioneering work of Turing \cite{Turing:1952xq}.
In such an approach, we take the limit of the cells to be a continuum. This modeling approach is very amenable to analytical work, and was for years the only approach used.
However, complexities arise when introducing cell division and inhomogeneities in the system.
As an alternative, with the rise of computational power, it has become possible to simulate cells as individual entities following simple rules.
This approach has the benefit that it naturally accounts for the individuality of cells, describing complex intrinsic properties flexibly.
The main drawback of this simulation framework is that it is hard to reach an analytical understanding that goes beyond specific parameter values.
Careful considerations are required to avoid over-complications of the model and extract reliable conclusions from their use.
The two techniques are not exclusive and can be complementary.
For example, Manukyan et al \cite{manukyan2017living} use a CM to explain the patterning of reptile scales, while proving that the approach maps to an ABM depending on the length scale at play.
The use of one method or the other will depend on the modeling purpose, with the question that we want to assess fitting naturally one or the other.

\begin{figure}[htb]
	\centering
	\centerline{\includegraphics[scale=.45]{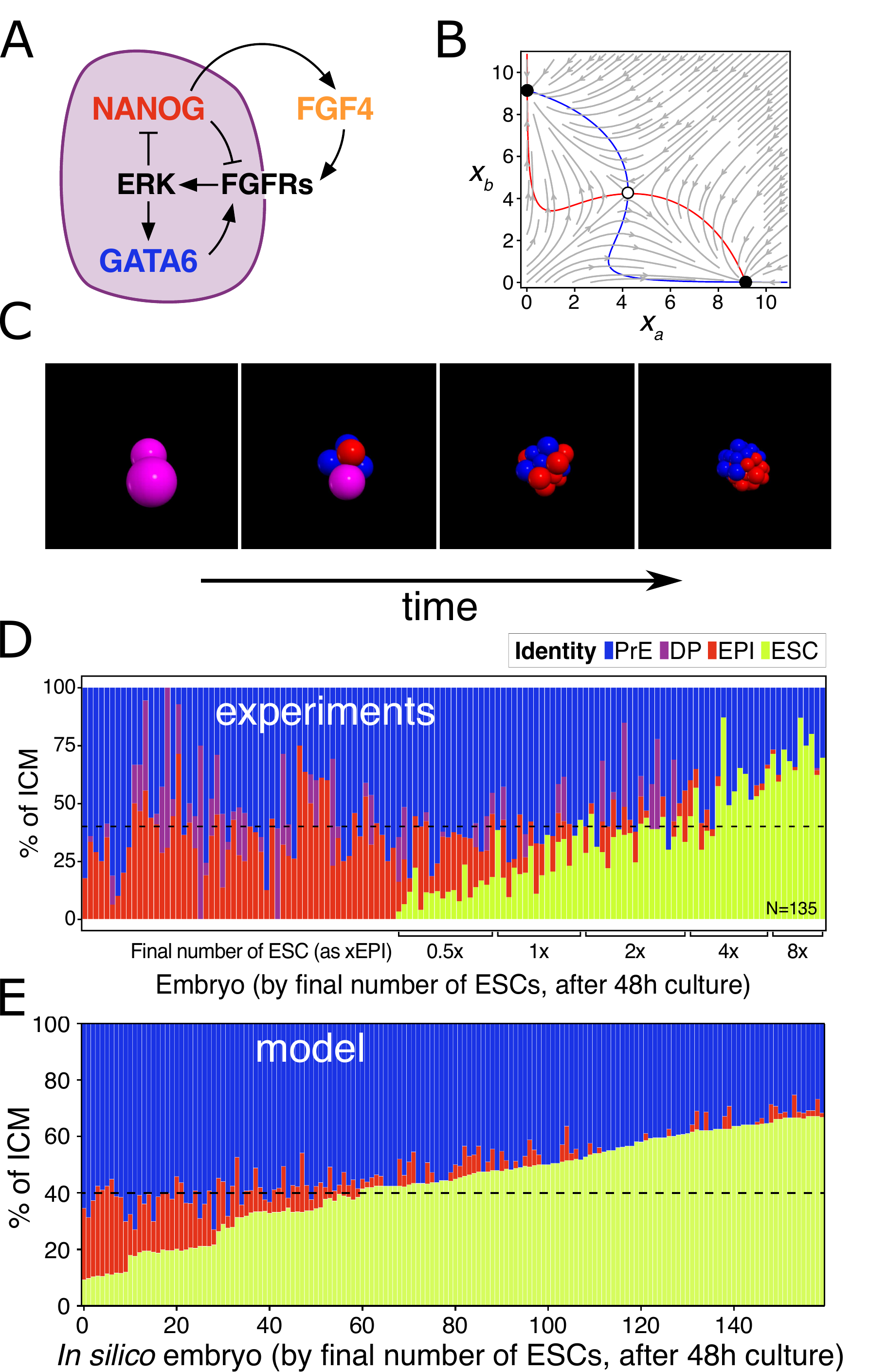}}
	\caption{Using a minimal model to study the robustness of collective cell-fate transitions. 
	A: Biochemical network describing the interaction between the master regulators of the epiblast and primitive endoderm (Nanog and Gata6, red and blue, respectively) in the early mouse embryo.
	The two transcription factors mutually inhibit each other via FGF/ERK signaling.
	B: Phase plane analysis of a minimal one-dimensional version of the network depicted in A.
	The black circles depict the two fates into which all cells alternatively differentiate as the embryo develops.
	C: Filmstrip showing the results of an agent-based model incorporating the minimal biochemical network shown in A,B.
	D,E: Cell-fate distribution at the late blastocyst stage when embryonic stem cells (light green) are added in different stages of the blastocyst development, in both experiments (D) and model (E).
	Adapted from \cite{saiz2020growth}.}
	\label{fig:2}
\end{figure}

As an illustrative example of this combined approach, Saiz et al \cite{saiz2020growth} investigated the robustness of cell fate proportions during the early development of the mouse embryo.
The current knowledge of this process points to a complex interaction of different interconnected pathways combined with intercellular communication (Figure~\ref{fig:2}A).
Despite the complexity of the genetic network, the dimensionality of the biochemical model can be reduced substantially through adiabatic considerations, leading to a single degree of freedom per cell.
This simple biochemical model is enough to explain the balanced distribution of fates between epiblast and primitive endoderm in terms of only cell-cell communication via FGF/ERK signaling, which leads to an effect akin to lateral inhibition in neighboring cells without the need of intracellular mutual inhibition between the master regulators of the two cell fates (Figure~\ref{fig:2}B).
To include the effects of cell mechanics and proliferation, the authors used an ABM where the cells obey the minimal biochemical model described above (Figure~\ref{fig:2}C).
The model reproduced the experimentally observed robustness to perturbations in the initial embryo size, using both  scaling experiments in which half-sized and double-sized embryos were considered, and perturbation experiments in which the relative sizes of the two fates were varied, by either addition of embryonic stem cells (Figures~\ref{fig:2}D,E) or targeted removal of cells of one type or the other at different times.
The agreement between the model and the experiments sheds light into the collective nature of cell-fate decisions in the early embryo.

\subsection{The heterogeneity of life: towards an non-equilibrium statistical mechanics of cell-fate decisions}

As we have discussed in the preceding section, the cell-fate transition landscape changes between cells under the effect of external signals and cell-cell interactions.
But there is an additional detail left: the stochastic nature of biology.
Noise is ubiquitous in cells, from the sensing of external signals to gene expression \cite{paulsson2005models,Eldar:2010aa}.
The effects of noise at the single-cell level have been studied for more than two decades \cite{paulsson2005models,elowitz2002stochastic,huang2005cell,richard2016single,guillemin2019drugs}.
A standard way to approach this issue so far has been to include noise sources as part of the modeling approach, usually working on small gene and protein circuits described by low-dimensional stochastic dynamical systems.
However, a fundamental understanding of the large-scale interplay between the noise and the behavior of cell-fate decisions is still missing.
The unique character of single-cell experiments should help us reach deeper insights into this problem.
In particular, the recent deluge of data is challenging established concepts such as that of cell identity \cite{Clevers:2017aa,stumpf2019machine}, and provides unprecedented insight into the dynamics of cell-fate decisions in commitment \cite{richard2016single, mojtahedi2016cell, li2013quantifying} and reprogramming \cite{li2013quantifying,zhang2014stem}.

\begin{figure}[htb]
	\centering
	\centerline{\includegraphics[scale=.45]{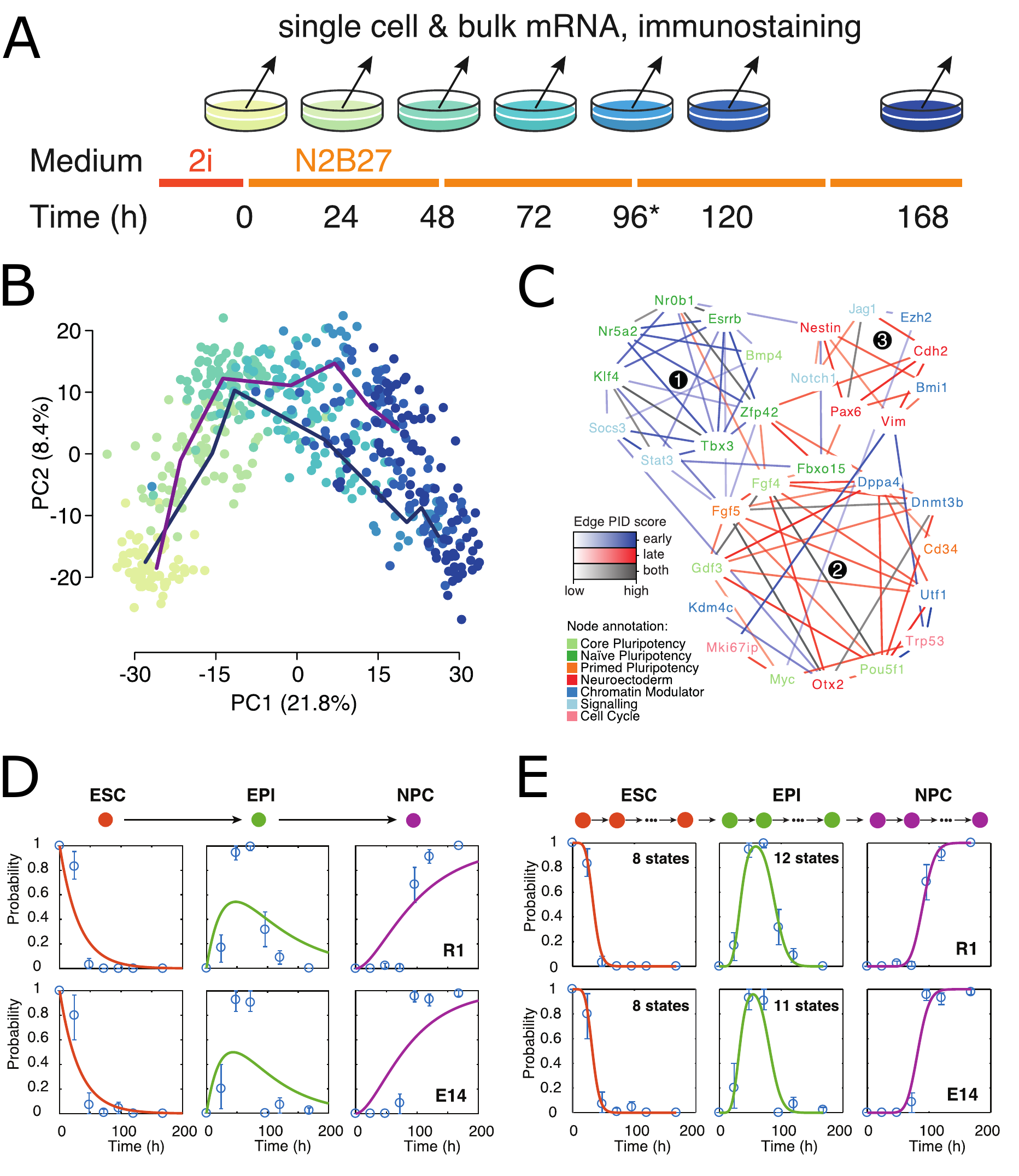}}
	\caption{Combining stochastic modeling and statistical data analysis to analyze cell-fate transitions.
	A: Experimental scheme for the \textit{in vitro} differentiation of stem cells to the neuroectoderm fate.
	B: Single-cell transcriptomics data projected in the plane of the first two principal components.
	The color code from yellow to blue represents increasing time.
	C: Gene regulatory network inferred from the single-cell transcriptomics data.
	D,E: Fitting the data to a stochastic first-order model without memory (D) and to a hidden Markov model (E).
	Adapted from \cite{stumpf2017stem}.}
	\label{fig:4}
\end{figure}

Previous proposals highlight the potential benefit of using the formalism of non-equilibrium statistical mechanics to account for this inherent stochasticity \cite{garcia2012towards, macarthur2013statistical}.
This route can nevertheless be challenging.
A step in this direction was taken by Stumpf et al \cite{stumpf2017stem}, who used single-cell data and a combination of machine learning and mechanistic modeling to examine the \textit{in vitro} differentiation of stem cells as they evolve towards the neuroectoderm fate (Figure~\ref{fig:4}A).
Cell trajectories towards the committed state, as inferred from single-cell transcriptome analysis, are highly stochastic (Figure~\ref{fig:4}B).
The single-cell data also allowed the authors to infer a regulatory network whose components change along the developmental path, activating and deactivating three different regulatory modules (Figure~\ref{fig:4}C, see labels 1-2-3 in the plot).
According to this analysis, cells evolve through three macrostates, which contain several microstate transitions.
A careful comparison between different alternative models reveals that the transitions are not consistent with Markovian  dynamics (Figure~\ref{fig:4}D), but they rather correspond to a non-Markovian stochastic process (Figure~\ref{fig:4}E).
This combination of stochastic modeling and statistical data analysis shows a way forward in our quest to understanding cell-fate transitions in development using high-throughput single-cell experimental methods.

\section{Final remarks}

Single-cell methods, particularly those that offer spatial resolution \cite{vickovic2019high,rodriques2019slide,eng2019transcriptome}, will be of great relevance to disentangle many open questions on the dynamics of the cell-fate decision landscapes.
However, we need to be mindful of the most effective way in which we can leverage those techniques to advance our understanding of the fundamental mechanisms behind multicellular self-organization.
Looking for the most minimalistic approach to explain current knowledge is at the heart of any modeling approach.
However, although simplicity is readily appreciated \cite{corson2017gene, corson2017self, saiz2020growth}, the continuous increase of detailed knowledge and the rising amount of data generated at single-cell resolution makes it very tempting to construct over-complicated models, which might not be the best way forward.

The reasons for keeping things simple are twofold.
First, over-detailed attitudes may obscure our understanding and curtail our ability to both search for trustful explanatory mechanisms and open up further avenues of research.
Second, despite the amount of data that single-cell techniques provide for us, we should always assess their explanatory capacity;
overwhelmingly complex models can be impossible to validate even with extensive experimental efforts.
Many studies have focused on the assessment of both issues, both looking systematically for simpler models compatible with the data \cite{proulx2017untangling,veleslavov2020repeated,camacho2021quantifying,babtie2014topological,saez2021quantitative} and assessing the limits of the information that we can obtain from those models and from the available data \cite{mattingly2018maximizing}.
These are the approaches that will generate understanding from the current revolution in data gathering.

\section{Acknowledgments}

This work was supported by the Spanish Ministry of Science, Innovation and Universities and FEDER (under projects PGC2018-101251-B-I00 and FIS2017-92551-EXP, and the ``Maria de Maeztu'' Programme for Units of Excellence in R\&D, grant CEX2018-000792-M), and by the Generalitat de Catalunya (ICREA Academia programme and grant 2017 SGR 1054).
G.T. is funded by PhD grant FPU18/05091 from the Spanish Ministry of Science, Innovation and Universities.

\end{document}